\documentclass[12pt]{article}
\usepackage[english]{babel}
\usepackage[normalem]{ulem}
\usepackage{multirow}
\usepackage{booktabs}
\usepackage{float}
\usepackage[text={15.5cm,22.5cm}]{geometry}
\usepackage{graphicx}
\usepackage[dvipsnames]{xcolor}
\usepackage{pdfcolmk}
\usepackage{amsmath}
\usepackage{amssymb}
\usepackage{slashed}
\usepackage{hyperref}
\usepackage{slashed}

\usepackage{tabularx}
\usepackage{cite}
\allowdisplaybreaks

\newcommand{\DM}{\mathrm{DM}}

\begin{document}

\title{\vspace{-0.8cm}
{\normalsize
\flushright TUM-HEP 1088/17\\ SI-HEP-2017-17, QFET-2017-15\\ LMU-ASC 44/17\\ MPP-2017-154\\}
\vspace{0.4cm}
\bf Sharp spectral features from light dark matter decay via gravity portals 
 \\ [6mm]}

\author{Oscar Cat\`a$^{1,2}$, Alejandro Ibarra$^{3,4}$, Sebastian Ingenh\"utt$^{3,5}$\\[2mm]
{\normalsize\it $^1$ Theoretische Physik 1, Universit\"at Siegen,}\\[-0.05cm]
{\normalsize\it Walter-Flex-Stra\ss e 3, D-57068 Siegen, Germany}\\[2mm]
{\normalsize\it $^2$ Ludwig-Maximilians-Universit\"at M\"unchen, Fakult\"at f\"ur Physik,}\\[-0.05cm]
{\normalsize\it Arnold Sommerfeld Center for Theoretical Physics, 80333 M\"unchen, Germany}\\[2mm]
{\normalsize\it $^3$ Physik-Department T30d, Technische Universit\"at M\"unchen,}\\[-0.05cm]
{\it\normalsize James-Franck-Stra\ss{}e, 85748 Garching, Germany}\\[2mm]
{\normalsize\it $^4$ School of Physics, Korea Institute for Advanced Study, Seoul 02455, South Korea}\\[2mm]
{\normalsize \it $^5$ Max-Planck-Institut f\"ur Physik (Werner-Heisenberg-Institut),}\\ [-0.05cm]
{\normalsize \it F\"ohringer Ring 6, 80805 M\"unchen, Germany}
}

\date{}
\maketitle
\thispagestyle{empty}
\vskip 1.5cm
\begin{abstract}
We investigate the phenomenology of dark matter decay assuming that it is induced by non-minimal coupling to gravity, when the dark matter mass is in the sub-GeV range, {\it i.e.} below the QCD confinement scale. We show that the decay of the singlet scalar dark matter candidate produces sharp features in the photon spectrum, in the form of lines, boxes, and also in the form of a novel spectral feature, characterized by the decay into $e^+e^- \gamma$ through a contact interaction, with decay branching fractions depending only on a single parameter, namely the dark matter mass.  We also derive upper limits on the strength of the gravity portal from the non-observation of sharp features in the isotropic diffuse gamma-ray spectra measured by COMPTEL, EGRET and Fermi-LAT, and the X-ray spectrum measured by INTEGRAL. Finally, we briefly comment on the impact of dark matter decay via non-minimal coupling to gravity on the reionization history of the Universe.
\end{abstract}
\newpage
\section{Introduction}
\label{sec:intro}

Cosmological observations reveal that approximately 16\% of the matter density of our Universe is in the form of protons, while the remaining 84\% is attributed to a non-luminous component~\cite{Ade:2015xua}, dubbed dark matter, possibly constituted by new particles not contained in the Standard Model (for reviews, see~\cite{Bertone:2010zza,Bergstrom:2000pn,Jungman:1995df,Bertone:2004pz}). Dark matter particles were produced in the very early stages of our Universe, therefore their presence today in galaxies, clusters of galaxies, and in the Universe at large scale requires their lifetime to be at least as long as the age of the Universe. 

A common strategy to make dark matter stable consists in postulating a new global symmetry, unbroken in the electroweak vacuum, under which the dark matter particle is charged while all particles in the observable sector remain neutral. This global symmetry is purported to forbid all operators leading to dark matter decay, rendering it absolutely stable. This discussion normally ignores the effects of gravity. However, it has long been argued that global symmetries cannot be preserved in the presence of gravitational interactions~\cite{Abbott:1989jw,Kallosh:1995hi,Hawking:1974sw}. Since curved spacetime is the natural arena where dark matter models should be embedded, this generates a mechanism for dark matter decay into the Standard Model particles via a gravitational portal. 

In~\cite{Cata:2016dsg,Cata:2016epa}, gravitationally-induced dark matter decay was explored for a class of operators linear in the dark matter field and coupled non-minimally to gravity through the Ricci scalar. In this framework, dark matter decays into Standard Model particles with a total decay rate suppressed by inverse powers of the Planck mass and with branching ratios which only depend on the dark matter mass. Refs.~\cite{Cata:2016dsg,Cata:2016epa} covered the range with dark matter masses larger than $\sim 1$ GeV, such that the Standard Model degrees of freedom (quarks, leptons, gauge bosons and the Higgs) provide a good description of the possible decay products. In contrast, for dark matter masses in the sub-GeV regime, the relevant degrees for freedom are the photon, the three neutrinos, the electron, the muon (and their antiparticles) from the electroweak sector, as well as the pions from the confinement of light quarks and gluons. The interactions of the latter are described by chiral perturbation theory, the theory of the strong interactions below the confinement scale  (for reviews, see~\cite{Ecker:1994gg,Leutwyler:1993iq}). Hence, the phenomenology of dark matter decay in the sub-GeV regime is qualitatively different to the one in the supra-GeV regime and a dedicated analysis is in order.

In this paper, we identify the dominant decay channels of a scalar singlet dark matter candidate with sub-GeV mass via the gravity portal, and we calculate the corresponding rates and branching ratios. Notably, we find that, for the whole mass range under consideration, the channels producing sharp spectral features in the photon spectrum have sizable branching fractions, thus providing a strong test on the size of the non-minimal couplings of dark matter to gravity.

\section{Gravity portal for light scalar dark matter decay}
\label{sec:2}

In the absence of gravity effects, the total action for the observable and dark sector can be written as
\begin{equation} \label{eq:CAJ}
\mathcal S = \int d^4x \left[\mathcal L^{\rm eff}_{\rm obs}(X) + \mathcal L_{\DM}(\phi,X)\right]\;,
\end{equation}
where $\phi$ is the dark matter field and $X$ generically denotes the dynamical degrees of freedom in the observable sector at the energy scale relevant for dark matter decay. $\mathcal L^{\rm eff}_{\rm obs}(X)$ is the effective Lagrangian of the observable sector and $\mathcal L_{\DM}(\phi,X)$ contains terms involving the dark matter field together with its possible interactions with the observable sector.
We also assume, as commonly done in the literature, the existence of a stabilizing global symmetry, under which the fields of the observable sector transform trivially but the dark matter field does not. 

In the presence of gravitational interactions, the stabilizing symmetry remains unbroken provided the dark matter only couples minimally to gravity. Nevertheless, non-minimal coupling to gravity may break the global symmetry and therefore induce dark matter decay. The dominant operators will be the ones with the lowest dimension. In this work we focus on the singlet scalar dark matter candidate which, as discussed in~\cite{Cata:2016dsg,Cata:2016epa}, can be non-minimally coupled to gravity through a gauge and Lorentz invariant operator in the action already at mass dimension 3,\footnote{Decay operators through non-minimal coupling to gravity for dark matter candidates with gauge charges and/or higher spin require higher dimensional operators in the action. As a result, these candidates are predicted to have naturally cosmologically long lifetimes, unless the dark matter mass is very large~\cite{Cata:2016epa}.} namely:
\begin{equation} \label{eq:CompleteActionJordan}
  \mathcal S = \int d^4x \sqrt{-g}\left[ -\frac{R }{2 \kappa^2}+ \mathcal L^{\rm eff}_{\rm obs}(X) + \mathcal L_{\DM}(\phi,X) -\xi M R\phi\right]\;,
\end{equation}
where $g$ is the determinant of the metric tensor $g_{\mu\nu}$, $\kappa=M_P^{-1}=\sqrt{8 \pi G}$ is the inverse (reduced) Planck mass, $M$ is a mass scale and $\xi$ is a dimensionless coupling. 

For dark matter heavier than the electroweak scale, $\mathcal L^{\rm eff}_{\rm obs}(X)$ conservatively corresponds to the Standard Model Lagrangian (although it may be extended to account for the dynamics of new, still undiscovered, degrees of freedom). In contrast, for dark matter lighter than the GeV scale, which is the focus of this paper, $\mathcal L^{\rm eff}_{\rm obs}(X)$ consists of terms describing the dynamics and interactions of the light degrees of freedom (the three pions, the photon, as well as the electron, the muon and the three neutrinos and their antiparticles)\footnote{Strictly, kaons and eta mesons are also dynamical at energy scales smaller than $\sim 1\,{\rm GeV}$. We will however not consider them explicitly in $\mathcal L^{\rm eff}_{\rm obs}(X)$, since the decays of a scalar dark matter particle with mass below $\sim 1\,{\rm GeV}$ into $KK$ or $\eta\eta$ pairs is kinematically forbidden.}, augmented with terms resulting from integrating out heavy particles from the action, which are off-shell at the energy scale relevant for dark matter decay (heavy fermions, $W$, $Z$ and Higgs). The effect of these heavy fields at low energies is encapsulated in the Wilson coefficients of four-fermion operators, as well as in the vacuum polarization of the photon. We will neglect the former, since they give rise to four-body decays, which are heavily suppressed by phase-space factors. Then, the effective Lagrangian relevant for our analysis can be cast as
\begin{align}\label{EFT2}
{\cal{L}}_{\mathrm{obs}}^{\mathrm{eff}}&=\sum_{\substack{f=e,\mu,\\\nu_1,\nu_2,\nu_3}}\left(\frac{i}{2}{\bar{f}}\mathop{\slashed{\nabla}}^{\longleftrightarrow}\!\!f-m_f{\bar{f}} f \right)\nonumber\\
&+\frac{f_{\pi}^2}{4}g^{\mu\nu}{\mathrm{Tr}}\left[D_{\mu}U^{\dagger}D_{\nu}U\right] +\frac{f_{\pi}^2m_{\pi}^2}{2}{\mathrm{Tr}}\left[ U^{\dagger}+U\right]\nonumber\\
& -\frac{1}{4}Z_3^{-1}g^{\mu\nu}g^{\lambda\rho}F_{\mu\lambda}F_{\nu\rho}\;.
\end{align}
In the first line above we have defined $\slash{\!\!\!\!\nabla}=\gamma^ae^{\mu}_a\nabla_{\mu}$, where $\gamma^a$ is a Dirac matrix, $e^{\mu}_a$ a vierbein and ${\nabla}_\mu=D_\mu -\frac{i}{4}e_\nu^b (\partial_\mu e^{\nu c}) \sigma_{bc}$, with $D_\mu$ the gauge covariant derivative. In the second line we have included the dominant operators of chiral perturbation theory.
The matrix  $U=\exp[i \vec \tau\cdot \vec \pi/f_\pi]$
contains the pion fields, $\pi^a$,  $a=1,2,3$ ($\tau^a$ are the Pauli matrices) and the covariant derivative is defined as $D_{\mu}U=\partial_{\mu}U+ieA_{\mu}[Q,U]$, with $Q={\mathrm{diag}}(2/3,-1/3,-1/3)$; the traces are taken over the flavor indices and $f_\pi=93\,{\rm MeV}$ is the pion decay constant. Finally, in the third line, $Z_3$ is the photon wavefunction renormalization constant, which receives contributions from all electrically charged degrees of freedom that have been integrated out at the cut-off scale of the theory, which in our case is $\sim 1\,{\rm GeV}$. These include the top, bottom and charm quarks, the tau lepton, the $W$ boson, as well as all hadrons made of light quarks except for the pions, which are the only dynamical hadronic degrees of freedom below $\sim 1\,{\rm GeV}$.\footnote{In extensions of the Standard Model, effects of new charged particles should also be included in the calculation.} 

The hadronic contributions to the wavefunction renormalization constant involve scales which lie in the non-perturbative regime of the strong interactions, and are therefore difficult to estimate. These contributions are however expected to be modest, certainly not larger than the one from all other degrees of freedom combined. In this work we will neglect the contribution from the hadronic states to the wave function renormalization constant. In practice, this will translate into a theoretical uncertainty in the calculation of $Z_3$, which we expect to be at most of ${\cal O}(1)$.	
Under this assumption, we obtain
\begin{align}
Z_3^{-1}&\approx 1-\frac{e^2}{8\pi^2}\left(\sum_{i= t,b,c,\tau} b_i \log\frac{\Lambda}{m_i}+b_W \log\frac{\Lambda}{M_W}\right)\;,
\end{align}
where $b_t=b_c=-16/9$, $b_b=-4/9$, $b_\tau=-4/3$ and $b_W=+7$. We will see later on that current experimental bounds justify this strategy. 

Given the form of the non-minimal operator, the action in Eq.~(\ref{eq:CompleteActionJordan}) can be recast as
\begin{equation} \label{eq:CompleteActionJordanDM}
  \mathcal S = \int d^4x \sqrt{-g}\left[ -\frac{R }{2 \kappa^2}\Omega^2(\phi,X) + \mathcal L_{\rm obs}^{\rm eff} + \mathcal L_{\DM} \right]\;,
\end{equation}
where 
\begin{equation}\label{eq:def-Omega}
\Omega^2(\phi,X) =1 + 2 \kappa^2 \xi M\phi\;.
\end{equation}
Written in this form, it is clear that the non-minimal operator modifies the Einstein-Hilbert action. The pure gravitational sector can be brought to canonical form through the field rescaling (Weyl transformation)
\begin{equation}\label{eq:transfo}
  \widehat g_{\mu\nu} = \Omega^2(\phi,X) g_{\mu\nu}\;,
\end{equation}
upon which the action gets transformed into the Einstein-frame form:
\begin{equation} \label{eq:CompleteActionEinstein}
  \mathcal S=\int d^4x \sqrt{- \widehat g}\left[ -\frac{\widehat R}{2 \kappa^2} + \frac{3}{\kappa^2}\widehat{g}_{\mu\nu}\frac{\widehat \nabla^\mu \Omega \widehat \nabla^\nu \Omega}{\Omega^2} + \mathcal {\widehat{L}}_{\rm obs}^{\rm eff} +\mathcal {\widehat{L}}_{\DM} \right]\;.
\end{equation}
The main benefit of this field redefinition is that dark matter interactions with the light particles can now be read out in a straightforward manner. The transformation of Eq.~(\ref{eq:transfo}) brings the effective Lagrangian of the observable sector to the form:
\begin{align} \label{eq:SMEinstein}
{\mathcal{\widehat{L}}}_{\rm obs}^{\mathrm{eff}} &=\sum_{\substack{f=e,\mu,\\\nu_1,\nu_2,\nu_3}}\left(\frac{i}{2 \Omega^3}{\bar{f}}\mathop{\widehat{\slashed{\nabla}}}^{\longleftrightarrow}\!\!f-\frac{m_f}{\Omega^4}{\bar{f}} f \right)\nonumber\\
&+\frac{f_{\pi}^2}{4\Omega^2}{\widehat g}^{\mu\nu}{\mathrm{Tr}}\left[D_{\mu}U^{\dagger}D_{\nu}U\right] +\frac{f_{\pi}^2m_{\pi}^2}{2\Omega^4}{\mathrm{Tr}}\left[ U^{\dagger}+U\right]\nonumber\\
& -\frac{1}{4}{\widehat Z}_3^{-1}{\widehat g}^{\mu\nu}{\widehat g}^{\lambda\rho}F_{\mu\lambda}F_{\nu\rho}\;,
\end{align}
where hatted quantities denote that they are expressed in the Einstein frame. 

Note that the Weyl transformation modifies not just the metric and the coefficients of the operators, but also the photon wavefunction renormalization constant. This can be understood as a consequence of the rescaling that the kinetic and mass terms of the integrated-out particles also undergo when transforming the action from the Jordan to the Einstein frame.\footnote{Inside loops only the propagators are relevant, so our discussion will be restricted to the free fields.} Since the masses of the particles that have been integrated out are much larger than the momentum of the dark matter particle, the field $\phi$ is slowly-varying and effectively behaves as a constant. In this limit, the Lagrangian of the integrated-out $W$ boson in the Einstein frame,
\begin{align}
\widehat{\cal L}_W= -\frac{1}{4}{\widehat g}^{\mu\nu}{\widehat g}^{\lambda\rho}W_{\mu\lambda}W_{\nu\rho}+\frac{1}{2}\frac{M_W^2}{\Omega^2} {\widehat g}^{\mu\nu} W_\mu W_\nu+\dots\;,
\end{align}
can be understood as having a $\phi$-dependent rescaled mass. Integrated-out fermions show a similar rescaling:
\begin{align}
\widehat {\cal L}_f &= \frac{i}{2\Omega^3}{\bar{f}}\mathop{\widehat{\slashed{\nabla}}}^{\longleftrightarrow}\!\!f-\frac{m_f}{\Omega^4}{\bar{f}} f +.\,.\,. \simeq 
\frac{i}{2}{\bar{\widehat{f}}}\mathop{\widehat{\slashed{\nabla}}}^{\longleftrightarrow}\!\!\widehat{f}-\frac{m_f}{\Omega}{\bar{\widehat f}} \widehat{f} +\dots\;,
\end{align}
which can be transferred entirely to the mass term once the field is canonically normalized (last equality). The effect of the Weyl transformation therefore amounts to a $\phi$-dependent change in the masses of the particles entering the wavefunction renormalization constant, which then reads:
\begin{align}
\widehat Z_3^{-1}(m_f, M_W)=Z_3^{-1}\left(\frac{m_f}{\Omega}, \frac{M_W}{\Omega}\right)\approx 1-\frac{e^2}{8\pi^2}\left(\sum_{i= t,b,c,\tau} b_i \log\frac{\Lambda\,\Omega}{m_i}+b_W \log\frac{\Lambda\,\Omega}{M_W}\right)\;.
\end{align}
Finally, upon expanding $\Omega$, one finds effective interaction terms between the dark matter field and the operator $F^{\mu\nu}F_{\mu\nu}$. This procedure is similar in spirit to the one presented in \cite{Shifman:1979eb} to determine $h\rightarrow \gamma\gamma$ using low-energy theorems.

It is important to note that in the Einstein frame the field $\phi$ is not canonically normalized, as apparent from Eq.~(\ref{eq:CompleteActionEinstein}). To bring the kinetic term of the dark matter field into the canonical form, we introduce the field  $\widehat{\phi}$, defined from the condition that
\begin{align}
{\cal L}_{\rm DM, kin} = \frac{1}{2} \left(\frac{1}{\Omega^2} +\frac{6 (\partial_\phi \Omega)^2}{\kappa^2 \Omega^2} \right) \widehat g^{\mu\nu} (\partial_\mu\phi)(\partial_\nu \phi)=
\frac{1}{2}\widehat g^{\mu\nu} (\partial_\mu\widehat{\phi})(\partial_\nu \widehat{\phi})\;.
\end{align}
The field $\widehat{\phi}$ is then related to $\phi$ by the transformation
\begin{align}
\widehat{\phi}=\frac{\sqrt{6}}{\kappa}\left\{y-y_0 +\frac{1}{2}\log\left[\frac{(1-y)(1+y_0)}{(1+y)(1-y_0)}
\right]\right\}\;,
\label{eq:chi-def}
\end{align}
with
\begin{align}
y=\sqrt{1+\frac{1+2 \kappa^2 \xi M \phi}{6  \kappa^2\xi^2 M^2}}\;,~~~~
y_0=\sqrt{1+\frac{1}{6  \kappa^2 \xi^2 M^2}}\;.~~~~
\label{eq:y-def}
\end{align}
Notice that when the dimensionless combination $\kappa \xi M\ll 1$, the effects of the canonical normalization can be safely neglected. However, as we will see below, these effects can have a significant impact on the phenomenology when $\kappa \xi M\gtrsim 1$. 

The terms of the Lagrangian inducing dark matter decay can thus be identified by expanding Eq.~(\ref{eq:SMEinstein}) in powers of $\widehat{\phi}$, and keeping the linear terms. Using
Eqs.(\ref{eq:chi-def},\ref{eq:y-def}), we obtain that for moderate values of $\kappa \xi M$ (at the energies we are working $\kappa\widehat{\phi} \ll 1$ is always fulfilled), the Weyl scaling factor can be approximated by:
\begin{align}
\frac{1}{\Omega^2}\simeq 1-\frac{2\kappa^2 \xi M \widehat{\phi}}{\sqrt{1+6 \kappa^2 \xi^2 M^2}}\;.
\end{align}
Furthermore, since we are interested in the dominant decay modes, we will expand the chiral field $U$ to quadratic order in the pion fields. The part of the effective Lagrangian linear in the (canonically normalized) dark matter field $\widehat{\phi}$ reads:
\begin{align} 
{\mathcal{\widehat{L}}}_{\rm obs}^{\rm eff}\supset -\frac{2\kappa^2 \xi M \widehat{\phi}}{\sqrt{1+6 \kappa^2 \xi^2 M^2}}
&\Big[\sum_{\substack{f=e,\mu,\\\nu_1,\nu_2,\nu_3}} \left(\frac{3i}{2}{\bar{f}}\gamma^\mu\partial_\mu f-2 m_f{\bar{f}} f\right)-\sum_{f=e,\mu} \left(\frac{3e}{2}{\bar{f}}\gamma^\mu A_\mu f\right)\nonumber \\
& +\frac{1}{2}\partial_{\mu}\pi^a\partial^{\mu}\pi^a - \frac{1}{2}m_{\pi}^2\pi^a\pi^a \nonumber \\
& +c_{\gamma\gamma}
F_{\mu\nu}F^{\mu\nu}\Big]\;,
\label{eq:LSM_ScalarDM_Einstein}
\end{align}
with
\begin{align}
c_{\gamma\gamma}\approx-\frac{e^2}{8\pi^2}\left(\sum_{i= t,b,c,\tau} b_i +b_W\right)=\frac{5e^2}{24\pi^2}\;,
\end{align}
from where one can extract the  dark matter decay vertices into electrons, muons, neutrinos, pions and photons. 
It is interesting to note that for $\kappa \xi M\ll 1$ the strength of the decay vertex is proportional to $\kappa\xi M$, while for large values, it becomes approximately constant, due to the effect of the canonical normalization of the dark matter field. The phenomenological consequences of this behavior will be discussed in the next section. 

\section{Decay rates and observational signals}

From the effective Lagrangian expressed in the Einstein frame, Eq.~(\ref{eq:LSM_ScalarDM_Einstein}), the dark matter partial decay rates can be calculated. In what follows we will drop the hats from the fields for ease of notation, bearing in mind that all fields are now canonically normalized.

The rates for the decay modes with a fermion-antifermion pair or with pions in the final state can be straightforwardly calculated, the result being:
\begin{align}
\Gamma_{\bar{f} f}=&~\frac{m_\phi^3}{8\pi}~\frac{\kappa^4 \xi^2 M^2 }{1+6\kappa^2 \xi^2 M^2}\
 \,x_{\! f}(1-4x_{\!f})^{3/2}\;, \nonumber \\
\Gamma_{\bar{f} f \gamma}=&~\frac{\alpha m_\phi^3}{16 \pi^2} ~\frac{\kappa^4 \xi^2 M^2 }{1+6\kappa^2 \xi^2 M^2}~ g(x_{\!f})\;, \nonumber\\
\Gamma_{\pi^+\pi^-}=&~\frac{m_\phi^3}{16\pi}~\frac{\kappa^4 \xi^2 M^2 }{1+6\kappa^2 \xi^2 M^2}~\left(1+2x_{\pi^+} \right)^2\,\left(1-4 x_{\pi^+} \right)^{1/2}\;,\nonumber\\
\Gamma_{\pi^0\pi^0}=&~\frac{m_\phi^3}{32\pi}~\frac{\kappa^4 \xi^2 M^2 }{1+6\kappa^2 \xi^2 M^2}~\left(1+2x_\pi \right)^2\,\left(1-4 x_\pi \right)^{1/2} \;,
\label{eq:rates}
\end{align}
where
\begin{align}
g(x)\equiv \Big(1+2x+24x^2\Big)\sqrt{1-4 x} 
&-12 x^2 \left(3-4 x\right) \log \Big( \frac{1-2 x +\sqrt{1- 4 x} }{2 x} \Big)\;,
\end{align}
and  $x_a=m_a^2/m_{\phi}^2$. 

The rate for the decay into two photons requires a more careful analysis, since  at order $e^2$ it receives contributions from the effective vertex $\widehat\phi F_{\mu\nu} F^{\mu\nu}$ in Eq.~(\ref{eq:LSM_ScalarDM_Einstein}), as well as from loops of electrons, muons and pions. The result is:
\begin{align}\label{gammagamma}
\Gamma_{\gamma\gamma}=&~\frac{ m_\phi^3}{16 \pi}\,\frac{\kappa^4 \xi^2 M^2 }{1+6\kappa^2 \xi^2 M^2} \, \left|  F_\ell \left(\frac{m_\phi^2}{4m_e^2}\right) + F_\ell \left(\frac{m_\phi^2}{4m_\mu^2}\right) + F_{\rm pion}
+c_{\gamma\gamma}\right|^2,
\end{align}
where $F_\ell(x)$ is the leptonic form factor, given by
\begin{align}
F_\ell(x)&=  \frac{e^2}{8\pi^2} \left[x + (x -1)f(x)\right]x^{-2}\;,
\end{align}
with
\begin{align}
f(x)= \begin{cases}
\arcsin ^2\sqrt{x} &\quad x\leq1, \\
\displaystyle{-\frac{1}{4}\left[\log \frac{1+\sqrt{1-x^{-1}}}{1-\sqrt{1-x^{-1}}} -i\pi\right]^2} &\quad x>1.
\end{cases}\;,
\end{align}
and $F_{\rm pion}$ is the pionic form factor, which will be neglected along with the hadronic contribution to $c_{\gamma\gamma}$. Therefore, and in contrast to the values of the decay rates into $f\bar f$, $f\bar f \gamma$ and $\pi\pi$, given in Eq.~(\ref{eq:rates}), which are exact up to higher order corrections, the rate into $\gamma\gamma$ can be claimed to be accurate only up to an ${\cal O}(1)$ factor. This level of accuracy, on the other hand, will be sufficient for drawing our main conclusions.

The decay widths for each of these final states depend on the dark matter mass and on the value of the parameter $\kappa \xi M$. Clearly, for smaller and smaller  $\kappa \xi M$, the total rate decreases. However, for asymptotically large values, even for transplanckian scales $\kappa \xi M \gg 1 $, the total rate reaches a finite asymptotic value, which is a factor $7/6$ larger than the corresponding rate for $\kappa \xi M = 1$. Notably, we see that due to the effect of the non-minimal coupling on the dark matter kinetic term, the decay rate cannot be significantly enhanced by taking $\kappa \xi M \gg 1 $.

\begin{figure}[t!]
	\begin{center}
		\hspace{-0.75cm}
		\includegraphics[width=0.65\textwidth]{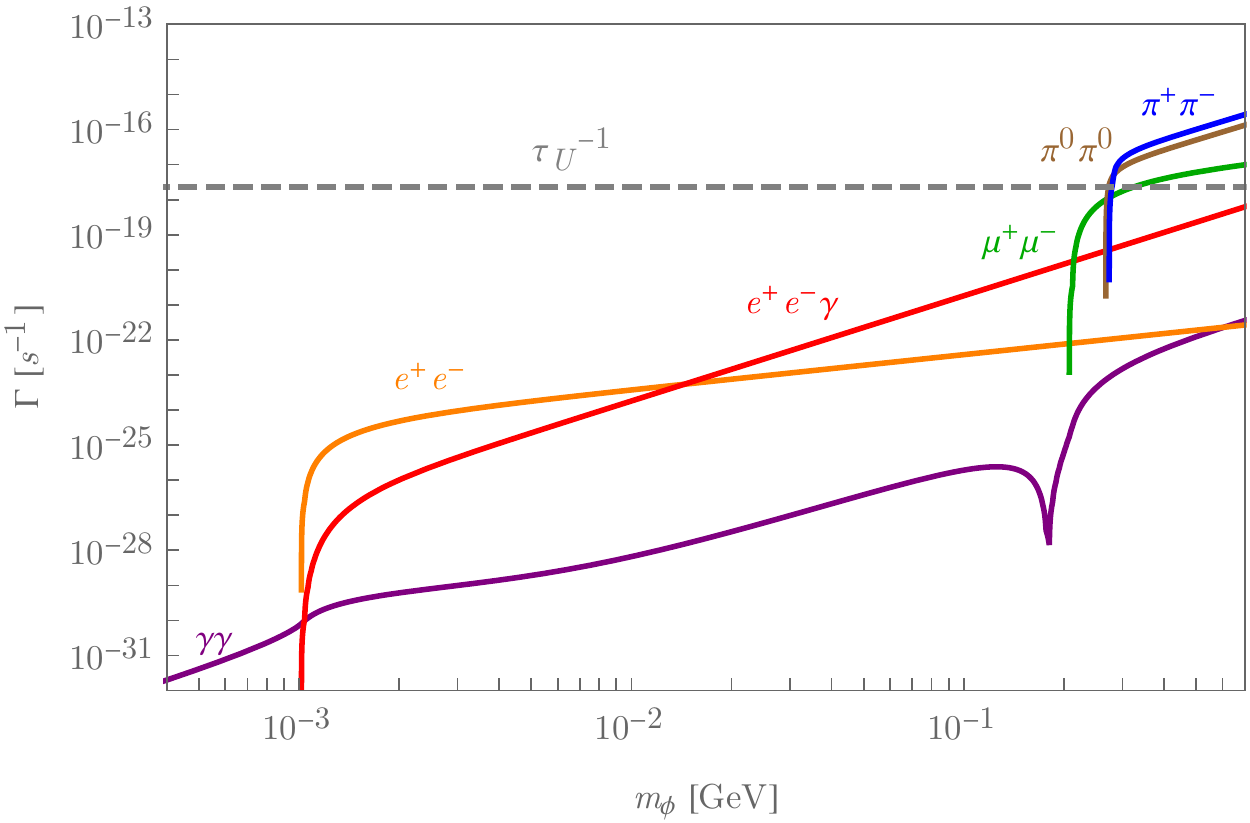}
	\end{center}
	\caption{Partial widths for the decay via non-minimal coupling to gravity of a singlet scalar dark matter candidate as a function of the mass, assuming $\kappa \xi M=1$.}
	\label{fig:BRs} 
\end{figure}

The inverse widths for the relevant channels are shown in Fig.\ref{fig:BRs}, for $m_\phi$ between 400 keV and 700 MeV, assuming $\kappa \xi M=1$. Above 700 MeV, chiral perturbation theory is no longer valid and our approach is not applicable. For dark matter lighter than $2 m_e$, only decays into photons or into neutrinos are kinematically accessible. For the latter, two-body decays are very suppressed by the conservation of angular momentum, hence we will only consider the former.\footnote{Other decays, such as  $2\nu2\bar\nu$, are suppressed by extra powers of the coupling constant and by the smaller phase space available in the decay, and can be safely neglected.} For dark matter masses between $2 m_e$ and $2 m_\mu$, the decay channels into $e^+e^-$ and $e^+e^-\gamma$ become kinematically accessible. Very close to the electron threshold, the decay into $e^+e^-$ dominates, as $\Gamma_{e^+e^-\gamma}$ is suppressed in this limit by an extra factor of $\alpha/\pi (1-4x_e)$. However, the rate for $\phi\rightarrow e^+e^-\gamma$ grows with the mass faster than the rate for $\phi\rightarrow e^+e^-$. We find that the branching fraction for $\phi\rightarrow e^+e^-\gamma$ equals 5\% for $m_\phi\simeq 3$ MeV, reaches 50\% for $m_\phi\gtrsim 14$ MeV and becomes the dominant decay channel until decays into muons become kinematically accessible, for $m_\phi=2m_\mu$. Close to the muon threshold, $\phi\rightarrow \mu^+\mu^-$ dominates over  $\phi\rightarrow e^+e^-$, due to the enhancement in the rate by the factor $m^2_\mu/m^2_e$, and over the decays $\phi\rightarrow \mu^+\mu^-\gamma,~e^+e^-\gamma$, due to the extra coupling constant and smaller phase space. The latter two processes have a larger rate than $\phi\rightarrow \mu^+\mu^-$ when $m_{\phi} \gtrsim 3$ GeV. However, for masses above the threshold of pion pair production, $m_\phi\gtrsim 270$ MeV, the decays $\phi\rightarrow \pi^+\pi^-,\pi^0\pi^0$ have rates larger than the processes with electrons or muons in the final state and become the dominant processes until $m_\phi\sim 700$ MeV, which is the largest mass considered in this work. 

The gray dashed line in the Figure indicates the age of the Universe, $\tau_U=4\times 10^{17}\,{\rm s}$. For dark matter masses above $\sim 270$ MeV, decays are so fast that the abundance of dark matter today would be much smaller than the observed value. In this regime, the parameter $\kappa \xi M$ is constrained to be smaller than 1, to ensure that the relic abundance of dark matter is compatible with present-day observations of our Universe. 

Stronger constraints on the parameter $\kappa\xi M$ can be derived from the non-observation of photon fluxes generated in the dark matter decay over the astrophysical background. The prompt photon flux is dominated by the decay channels $\gamma\gamma$, $f \bar f \gamma$ and $\pi^0 \pi^0$. The differential spectrum in each case reads:
\begin{align}
\frac{d N_\gamma^{(\gamma\gamma)}}{d y_\gamma}&=2 ~\delta\left(y_\gamma-\frac{1}{2} \right)\;, \nonumber \\
\frac{dN_\gamma^{(f\bar f \gamma)}}{d y_\gamma}&= \frac{24}{g(x_f)} \left( 1+ 2 x_f - 2 y_\gamma \right)  y_\gamma \sqrt{1-\frac{4 x_f}{1-2 y_\gamma}}\;,\nonumber \\
\frac{dN_\gamma^{(\pi^0\pi^0)}}{d y_\gamma}&=\frac{8}{\sqrt{1-4x_{\pi^0}}}\Theta(y_\gamma-y_-)\Theta(y_+-y_\gamma)\;,
\end{align}
with $y_\gamma\equiv E_\gamma/m_\phi$, $y_{\pm}\equiv (1\pm\sqrt{1-4x_{\pi^0}})/4$ and 
$\Theta(x)$ the Heaviside function.  

Notably, for most values of the dark matter mass, the decay via the gravity portal predicts a sharp  feature in the photon energy spectrum. For $m_\phi< 2m_e$, the dominant decay channel is $\phi\rightarrow \gamma\gamma$, which produces a line in the photon energy spectrum~\cite{Srednicki:1985sf,
Rudaz:1986db,Bergstrom:1988fp}. For $2 m_{\pi^0}< m_\phi\lesssim 1\,{\rm GeV}$, the dominant decay channel is $\phi\rightarrow \pi^0\pi^0$, which produces a gamma-ray box from the decay in flight of the pions into two photons~\cite{Ibarra:2012dw}. Finally, for $2 m_e\lesssim m_\phi < 2 m_{\pi^0}$, we find a new sharp spectral feature, which arises from the decay vertex $\phi \bar e \gamma^\mu A_\mu e$. 

\begin{figure}[t!]
	\begin{center}
		\hspace{-0.75cm}
		\includegraphics[width=0.7\textwidth]{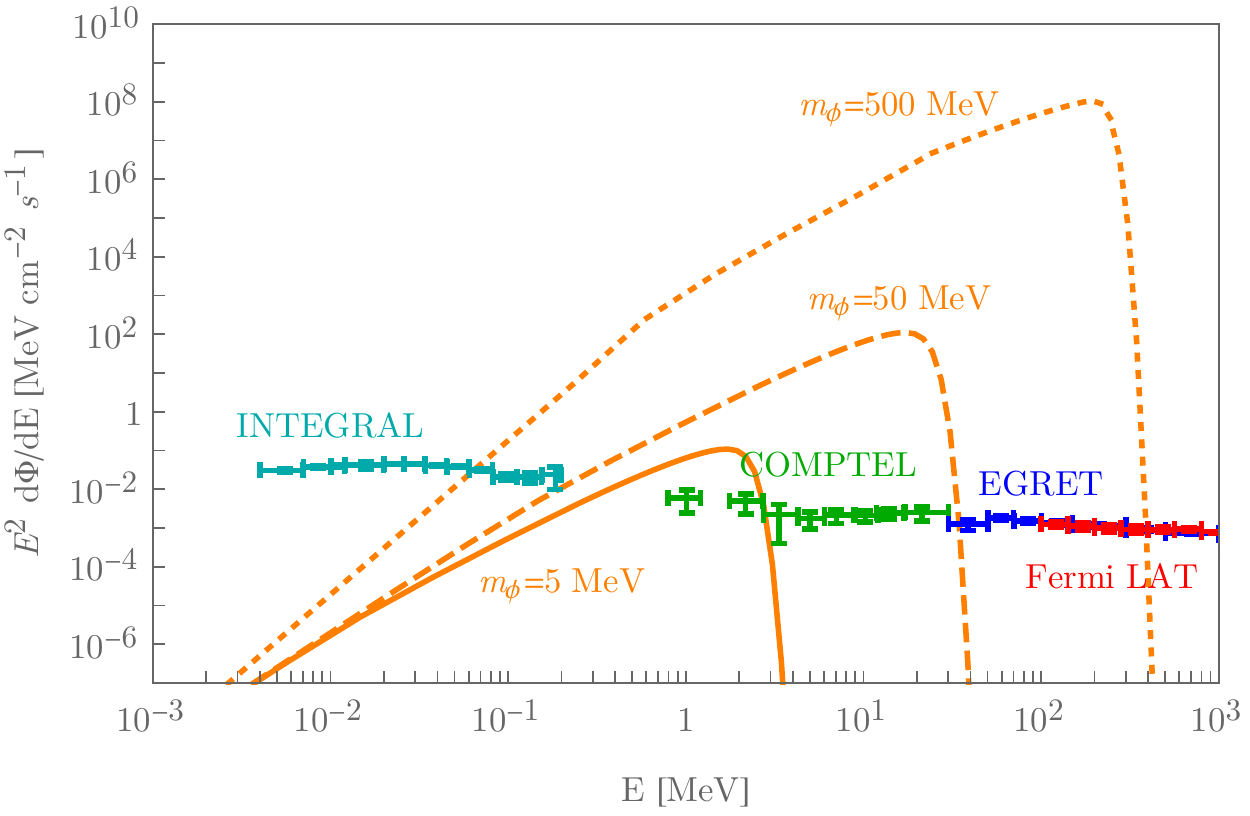}
	\end{center}
	\caption{Expected differential flux from the decay of a singlet scalar dark matter candidate with mass 5 MeV, 50 MeV and 500 MeV via non-minimal coupling to gravity, assuming $\kappa \xi M=1$.}
	\label{fig:spectra} 
\end{figure}

In Fig.~\ref{fig:spectra} we show the expected isotropic diffuse photon spectra from the decay of a dark matter particle with mass 5 MeV, 50 MeV and 500 MeV, assuming $\kappa \xi M=1$. The predicted flux includes contributions from the decay of the cosmological dark matter, with density $\Omega h^2\simeq 0.12$ as reported by {\it{Planck}}~\cite{Ade:2015xua}, and from the decay of Galactic dark matter, assumed to be distributed following a Navarro-Frenk-White~\cite{Navarro:1996gj} profile with scale factor equal to 24 kpc~\cite{Cirelli:2010xx}, normalized such that the dark matter density at the position of the Solar System is $\rho_{\rm loc}=0.3\,{\rm GeV}/{\rm cm}^3$. The detector response was simulated by adopting a fixed 10\% energy resolution over the whole range. The Figure also shows the isotropic diffuse X-ray spectrum in the energy range 5-100 keV, as determined by INTEGRAL~\cite{Churazov:2006bk}, and gamma-ray spectrum between 0.8 and 30 MeV, between 30 MeV and 50 GeV, and between 100 MeV and 820 GeV, as determined by  EGRET~\cite{Strong:2004ry}, COMPTEL~\cite{Weidenspointner} and Fermi-LAT \cite{Ackermann:2014usa}, respectively. 

As apparent from the Figure, when $\kappa \xi M=1$ the expected photon flux from dark matter decay can exceed the measured flux by many orders of magnitude. Furthermore, in all the cases, the energy spectrum presents a sharp fall-off close to the kinematic endpoint of the spectrum, which can be easily discriminated from the featureless astrophysical background. Therefore, the non-observation of sharp features in the isotropic diffuse photon spectrum translates into strong limits on the combination of parameters $\kappa \xi M$. 

\begin{figure}[t!]
	\begin{center}
		\hspace{-0.75cm}
		\includegraphics[width=.7\textwidth]{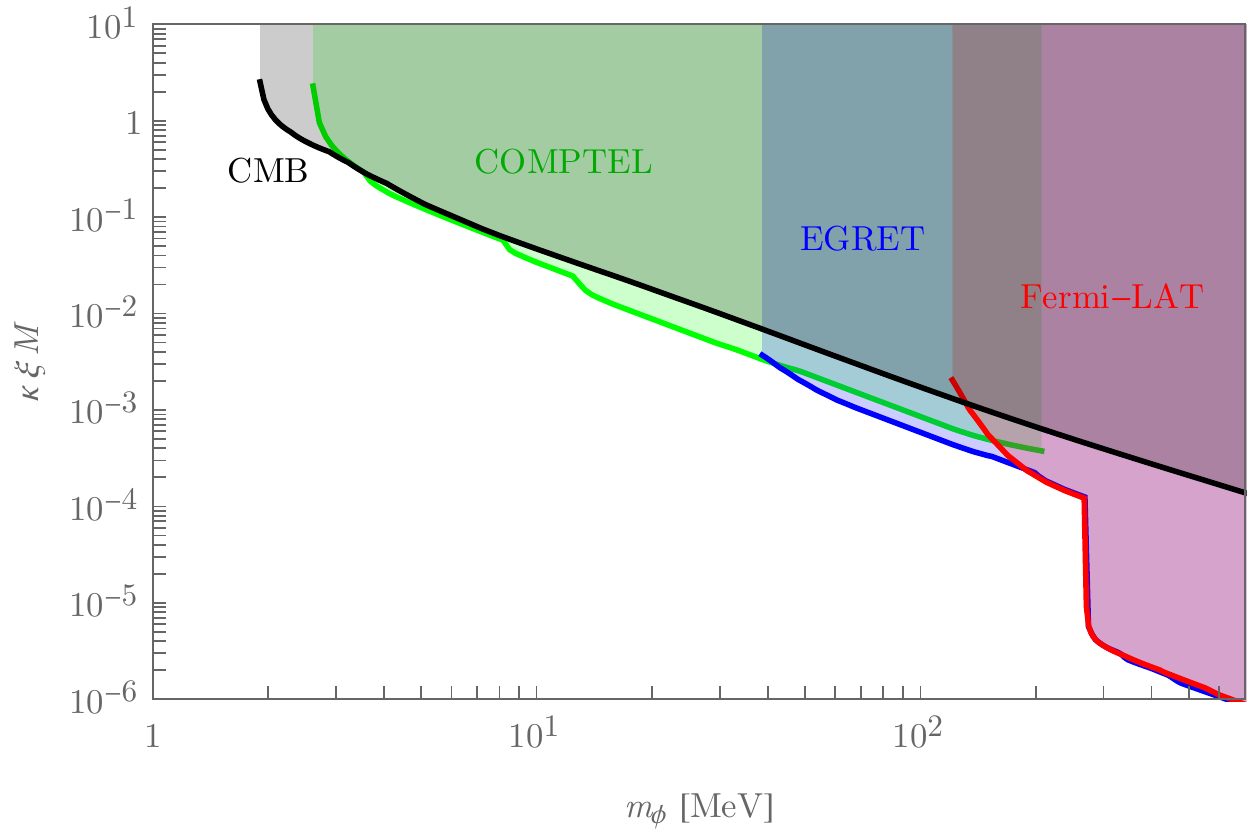}
	\end{center}
	\caption{Upper limit on $\kappa \xi M$ from the non-observation of sharp spectral features in the isotropic diffuse gamma-ray spectrum measured by COMPTEL (green line), EGRET (blue line) and Fermi-LAT (red line), as well as from requiring compatibility with CMB measurements at 95\% confidence level (black line).}
	\label{fig:xiLim} 
\end{figure}

In Fig.~\ref{fig:xiLim} we show conservative limits on $\kappa\xi M$, calculated from requiring that in any energy bin of one of the four experiments under consideration, the photon flux from dark matter decay only does not exceed the measured flux by more than $2\sigma$; for each experiment we used the appropriate energy resolution. We note that for very low dark matter mass the constraints disappear. The reason is that for every $m_\phi$ the decay rate has an upper limit, which is reached when $\kappa \xi M\gg 1$ ({\it cf.} Eq.~(\ref{eq:rates})). Therefore, this scenario will be unconstrained if the maximum flux predicted from dark matter decay is below the experimental sensitivity. We find numerically that this occurs for {\mbox{$m_\phi\lesssim 2$ MeV.}}

The observation of the spectral feature from dark matter decay via the gravity portal is subject to other phenomenological constraints, mainly from the modifications of the cosmic microwave background (CMB) anisotropy power spectrum induced by the injection of energy after recombination, which could significantly modify the ionization history of the Universe~\cite{Adams:1998nr,Chen:2003gz,Padmanabhan:2005es,Chluba:2011hw}. We have estimated the CMB constraints on our scenario using the \texttt{Project Epsilon} code \cite{Slatyer:2016qyl}; the resulting limits are shown as a black line in Fig.~\ref{fig:xiLim}. Interestingly, the CMB constraints are weaker than the ones from gamma-ray telescopes, thus opening the possibility of observing sharp spectral features from dark matter decay via the gravity portal in future missions, like e-ASTROGAM~\cite{DeAngelis:2016slk}.

\section{Conclusions}
\label{sec:conclusions}

Stabilizing global symmetries for dark matter do not need to be preserved by gravitational interactions. Therefore, in curved spacetime, which is the natural framework to describe dark matter, terms inducing dark matter decay through gravity are a logical possibility. In this paper we have investigated the impact of operators linear in the dark matter field and proportional to the Ricci scalar on the stability of a singlet scalar dark matter candidate with sub-GeV mass. In this energy regime, the final states kinematically accessible for dark matter decay contain pions, photons, electrons, muons and neutrinos (and their antiparticles). We have identified the dominant decay channels and calculated the corresponding decay widths in terms of known parameters of the Lagrangians of chiral perturbation theory and quantum electrodynamics. The only free parameters in our analysis are therefore the dark matter mass and the dark matter coupling strength to the Ricci scalar, given by $\kappa \xi M$, with $\xi$ ($M$) a dimensionless (mass dimension) parameter and $\kappa$ the inverse reduced Planck mass. 

We have shown that the decay via the gravity portal always produces a sharp feature in the isotropic diffuse photon spectrum, which could be readily discriminated from the featureless astrophysical backgrounds, thus providing a probe of the dark matter non-minimal coupling to gravity. Concretely, for $m_\phi< 2m_e$, the dominant decay channel is $\phi\rightarrow \gamma\gamma$, which produces a line in the photon energy spectrum; for $2 m_{\pi^0}< m_\phi\lesssim 1\,{\rm GeV}$, the dominant decay channel is $\phi\rightarrow \pi^0\pi^0$, which produces a gamma-ray box from the decay in flight of the pions into two photons; and for $2 m_e\lesssim m_\phi < 2 m_{\pi^0}$, we have found a new sharp spectral feature, which arises from the decay vertex $\phi \bar e \gamma^\mu A_\mu e$. The non-observation of sharp gamma-ray features in current experiments already places strong constraints on the dark matter non-minimal coupling to gravity, {\it{e.g.}} $\kappa\xi M\lesssim 10^{-6},\, 10^{-3}$ and $10^{-1}$ for $m_\phi\simeq 500,\, 50$ and $5\,{\rm MeV}$, respectively. Complementary limits on the parameter space stem from requiring that the decay products do not significantly modify the reionization history of the Universe. These limits are comparable or weaker than our conservative limits from the non-observation of sharp features in the photon spectrum. Planned X- and gamma-ray instruments will continue testing the possibility of dark matter decay via the gravity portal. 

\section*{Acknowledgements}
We are grateful to Fred Jegerlehner for useful discussions.
This work has been partially supported by the DFG cluster of excellence EXC 153 ``Origin and Structure of the Universe'' and by the Collaborative Research Center SFB1258.

\bibliographystyle{JHEP-mod}

\bibliography{references}

\end{document}